\documentclass[twocolumn,superscriptaddress,longbibliography,aps,pra,preprintnumbers]{revtex4-1}
\usepackage{hyperref}
\usepackage{epsfig}
\usepackage{graphicx}
\usepackage{epstopdf}
\usepackage{tikz}
\usepackage{float}
\usepackage{amsmath}
\usepackage{amssymb}
\usepackage{bm}
\usepackage[normalem]{ulem}

\begin{document}

\title{Influence of assisted hopping interaction on the linear conductance of quantum dot}

\author{G. G\'{o}rski}
\email{ggorski@ur.edu.pl}
\author{K. Kucab}
\email{kkucab@ur.edu.pl}
\affiliation{Faculty of Mathematics and Natural Sciences, University of Rzesz\'{o}w,
PL-35-310 Rzesz\'{o}w, Poland}

\begin{abstract}
We study the tunneling conductance and magnetization of a correlated quantum dot coupled to external metallic leads, considering the occupancy dependent hybridization. Using a modified equation of motion approach and self-consistent perturbation theory we show that the assisted hopping processes can be responsible for appearance of a plateau in the linear conductance. This process breaks the particle-hole symmetry and is manifested in conductance and magnetization. In presence of the external magnetic field the correlated hybridization enhances the spin polarization.
\end{abstract}

\maketitle
\section{Introduction}

Electronic transport through quantum dots (QD) or single electron transistors (SET) has recently been widely studied, both experimentally \cite{Kretinin-2011, Viel-2000, c50, c51, c52} and theoretically \cite{Meir-1993, Aligia-2006, c53, c54, c55, c56, c57}. It has a perspective for applications in modern electronics (spintronics) based on nanoscopic structures. These systems can be analyzed within the single impurity Anderson model (SIAM), which allows to investigate many interesting phenomena, e.g. the Coulomb blockade or the Kondo effect. Using the SIAM one can explain a narrow peak appearing at the Fermi energy in the density of states of the quantum dot, a zero-bias maximum in differential conductance and its splitting in a magnetic field.

In this work we study the influence of assisted hopping effect on the differential conductance. The assisted hopping term, $\Delta V_{k\alpha}$, is widely used to describe the properties of the bulk systems \cite{Hubbard-1963,c27,c28,c29, c30, c31, c32, c33, c34, c35, c36, c37, c38, c39, c40, c41, c42,Campbell-1990}. This interaction was used as the mechanism responsible for the hole high temperature superconductivity \cite{c27,c28}, the metal-insulator transition \cite{Hubbard-1963,c29,c30,c31}, the charge-density wave and the spin-density wave \cite{c32,Campbell-1990}. The assisted hopping interaction was also often used to describe the metallic ferromagnetism \cite{c33,c34,c35,c36,c37,c42}. The results presented in these articles  show strong enhancement of the ferromagnetic ordering by the term $\Delta V_{k\alpha}$. One special feature of this interaction is that even for the symmetrical density of states (DOS) it breaks the symmetry with respect to the half-filling carrier concentration point. Estimates for the assisted hopping interaction energy showed that this interaction is comparable to or larger than a magnitude of the direct hopping of real systems \cite{c48}.

The assisted hopping interaction was also used to describe the transport properties of the nanodevices \cite{Meir-2002, c43, c44, c45, c46, c47}. Meir and co-authors \cite{Meir-2002} proposed to use this interaction to explain the "0.7 anomaly" in the quantum point contacts linear conductance. This interaction also affects the off-diagonal pairing correlations in the Anderson impurity model \cite{c44}. It also strongly affects the thermoelectric effect \cite{c45}. Lin et al.~\cite{c47} used this interaction to describe the properties of molecular transistors, by means of the numerical renormalization group method. They showed that the correlated hybridization breaks the particle-hole symmetry for Kondo temperature and for the linear conductance. However, their results do not show the generation of "0.7 anomaly" in the conductance, as was postulated in \cite{Meir-2002}. This can be due to the use of strong Coulomb interaction ($U/\Gamma=10$). For the quantum point contacts the plateau was frequently observed, both for the temperature dependent \cite{c58, c59, c60} and the magnetic field dependent conductances \cite{c58, c61}. In this work we will show that for extended SIAM model, where the correlated hybridization is taken into account, one can obtain the conductance plateau both for the dependence on the temperature and on the magnetic field. In contrast to \cite{c47} we will use the weaker Coulomb interaction. The experimental data for a quantum dot shows the existence of conductance plateau for strongly coupled quantum dots \cite{Kretinin-2011, Viel-2000}.

This paper is organized as follows: in Section 2 we develop our approach to analyze the single impurity Anderson model. Using the modified EOM approach \cite{c49,Gorski-2015} and modified perturbation theory \cite{Aligia-2006, Potthoff-1997,  Domanski-2016, Kajueter-1996} we obtain expressions for the self-energy and the Green function in the presence of Coulomb repulsion. In this section we also show the influence of assisted hopping interaction on the quantum dot self-energy.  In Section 3 we present the numerical results based on our approach. At first we focus on the case without magnetic field. Calculating the dot Green function we obtain the spectral function for the quantum dot, the zero-bias linear conductance as a function of the dot energy and the differential conductance $dI/dV$  as a function of the bias voltage. The dependence of the linear conductance on the temperature is also analyzed. It is shown that the increase of temperature reduces the value of differential conductance. We also show that the assisted hopping interaction can cause the formation of linear conductance plateau. The value of $G_0$ for which the plateau occurs will depend on the Coulomb interaction and the assisted hopping interaction. The increase of both, the Coulomb and assisted hopping interaction causes the decrease of $G_0$ for which the plateau occurs. We show, that for strong correlation the plateau of linear conductance can be difficult to observe. In the second part of Section 3 we present the results for magnetization and  the linear conductance as a function of external magnetic field $\mathbf{B}$. Our results are compared with the experimental data and previous calculations. The summary and conclusions are given in Section 4.

\section{The model}

We consider the correlated QD coupled to two metallic leads. In our model we assume 
that the strength of this interaction will depend on the QD occupation described by 
the effective hopping operator $\hat V_{k\alpha\sigma }^{\rm{eff}}  = V_{k\alpha }  + \Delta V_{k\alpha } \hat n_{d - \sigma }$, 
where $V_{k\alpha }$  and $\Delta V_{k\alpha }$ is the coupling and assisted coupling between 
the  $\alpha$ lead and the dot, respectively. 
For further analysis we will assume that the assisted coupling 
is described by the assisted coupling parameter $\alpha _{V}$  
via the relation $\Delta V_{k\alpha} = \alpha_V V_{k\alpha}$. 
The Hamiltonian of our model has the form
 \begin{eqnarray}
H &=& \sum\limits_\sigma  {\varepsilon _{d\sigma } \hat n_{d\sigma } }  + U\hat n_{d \uparrow } \hat n_{d \downarrow }  + \sum_{\substack{k\sigma  \\ \alpha  = L,R}} {(\varepsilon _{k\alpha }  - \mu _\alpha  )\hat n_{k\alpha \sigma } }\nonumber \\
	 &+& \sum_{\substack{k\sigma  \\ \alpha  = L,R}} {\left[ {V_{k\alpha }^{} (1 + \alpha _V \hat n_{d - \sigma } )d_\sigma ^\dag  c_{k\alpha \sigma }  + h.c.} \right]}, 
\label{Hamiltonian}
\end{eqnarray}		
where $d_\sigma^\dag(d_\sigma)$ are the creation (annihilation) operators 
for the dot electron with spin $\sigma$, $c_{k\alpha\sigma}^\dag(c_{k\alpha \sigma })$  
are the creation (annihilation) operators for the conduction lead electron, $\alpha=L,R$ 
corresponds to the left and right leads, $\varepsilon _{k\alpha }$
 is the energy dispersion of $\alpha$ lead, $\mu_\alpha$ is the chemical potential 
of $\alpha$ lead, $U$ is the on-site Coulomb interaction between electrons on the dot 
and $\varepsilon _{d\sigma } $  is the dot energy. In external magnetic field the 
dot energy is split into two dot energies  $\varepsilon _{d\uparrow }=\varepsilon _{d}-B$  
 and $\varepsilon _{d\downarrow }=\varepsilon _{d}+B$  by the Zeeman energy $2B=g\mu_BH$.

We will calculate the impurity Green's function $G_{d\sigma}(\omega)=\langle\langle d_\sigma;d_\sigma^\dag\rangle\rangle_\omega $ 
 and its self-energy $\Sigma_{d\sigma } (\omega )$  using the modified equation of motion method \cite{c49,Gorski-2015}. 
In this method the two-time Green's function is differentiated over the first and the second time variable. 
As the result, after Fourier transform, we obtain the following (fully equivalent) equations of motion: 
\begin{eqnarray}
\omega\langle\langle A;B\rangle\rangle _\omega  \; = \;\langle [A,B]_ \pm  \rangle  + \;\langle \langle [A,H]_ -  ;B\rangle \rangle _\omega  
\label{EOM}
\end{eqnarray}		
and
\begin{eqnarray}
 - \omega \langle \langle A;B\rangle \rangle _\omega  \; =  - \;\langle [A,B]_ \pm  \rangle  + \;\langle \langle A;[B,H]_ -  \rangle \rangle _\omega,  
 \label{EOM2}
\end{eqnarray}
where $A,B$ are fermionic (bosonic) operators and $\pm$ denotes anticommutator and commutator, respectively. 

The equation of motion method (EOM) is a well-established method to
analyze the transport properties of quantum dot systems \cite{c53, c54, c56, Kashcheyevs-2006, Roermund-2010, Monreal-2005,Ochoa-2014,Levy-2013a,Levy-2013b}.
In a classical form of this method, to obtain the quantum dot Green's function one uses only the equation \eqref{EOM}. This method was used for equilibrium and non-equilibrium case. 
For uncorrelated quantum dot ($U=0$) and in the equilibrium case this equation gives the exact results, but for the Hamiltonian which includes $U$ interaction, the computation of single-particle Green's function generates additional higher-order Green functions. As the result, we obtain the infinite hierarchy of EOM of higher-order Green functions. Standard approximation for classical EOM method, based on the Lacroix decoupling scheme \cite{Lacroix-1981}, truncates the EOM hierarchy at second order in hybridization term.
The use of this approximation causes that the classical EOM method gives wrong results in the particle-hole symmetric case, because it causes the disappearance of the Kondo peak. As the result, in this case the linear conductance takes relatively small values and the unitary limit for the linear conductance is not reached.
The discussion of shortcomings of the classical EOM approach is presented by Levy and Rabani \cite{Levy-2013a}.

In order to reduce the disadvantages of the classical EOM method, there were made attempts to apply the approximation based on a truncation of the equations of motion at the fourth order in hybridization term \cite{Roermund-2010, Monreal-2005}, however, even at such a high order of Green functions, the results are not fully satisfying.

The modified equation of motion method (see \ref{sec.appendixA}) proposed by us is based on the use of the second form of equation \eqref{EOM2}
to obtain the higher-order Green's function. The equation of motion method with the differentiation over the first and the second time variable was introduced by Tserkovnikov \cite{Tserkovnikov-1962}. Additionally, we use the irreducible Green functions method \cite{Kuzemsky-2002} in which the approximations can be generated not by truncating the set of coupled equations of motion but by a specific approximation of the functional form of the higher-order irreducible Green functions. As the result, we obtain the $U^2$ order self-energy. For the modified equation of motion method the Kondo resonance peak is visible for both the particle-hole symmetric case ($ n_d =1$) and for the asymmetric case. For the particle-hole symmetric case the unitary limit for the linear conductance is satisfied. Moreover, the results for the linear conductance obtained by the use of this method are qualitatively similar to the results obtained by the numerical renormalization group
and quantum Monte Carlo calculations (see \cite{Gorski-2015}).

	Using Eqs. \eqref{EOM}, \eqref{EOM2} and the Hamiltonian \eqref{Hamiltonian} we obtain the expression for the quantum dot Green's function, $G_{d\sigma}(\omega)$, which has the following form \cite{Gorski-2015,Gorski-2016}:
\begin{eqnarray}
G_{d\sigma }^{} (\omega ) = \frac{1}{{\omega  - \varepsilon _{d\sigma }  - i\Gamma _\sigma ^{\rm{eff}} (\omega ) -\Sigma _{d\sigma } (\omega )}},
\label{Green}
\end{eqnarray}   
where the quantum dot self-energy 
\begin{eqnarray}
\Sigma _{d\sigma } (\omega )=Un_{d-\sigma }+ B_{\sigma} + \Sigma _{d\sigma }^{'} (\omega ) 
\label{Sigmad}
\end{eqnarray}  
is a sum of the Hartree-Fock part of Coulomb correlation $Un_{d-\sigma }$, the correlation parameter $B_{\sigma}$ related to the assisted hopping interaction and the higher order part of self-energy  $\Sigma _{d\sigma }^{'} (\omega )$.
The expression  $\Gamma_\sigma^{\rm{eff}} (\omega )$ is the effective hybridization function given by
 \begin{eqnarray}
\Gamma _\sigma ^{\rm{eff}} (\omega ) = {\rm Im} \sum_{\substack{k  \\ \alpha  = L,R}} {\frac{{\left( {V_{k\alpha \sigma }^{\rm{eff}} } \right)^2 }}{{\omega  - \varepsilon _{k\alpha } }}}, 
\label{Gamaeff0}
\end{eqnarray} 	
where we introduced the effective hybridization matrix element $V_{k\alpha \sigma }^{\rm{eff}}=V_{k\alpha } \left(1+\alpha_V n_{d-\sigma}\right)$. If we assume that the hybridization matrix element, $V_{k\alpha }$, is $\mathbf{k}$-independent, we obtain: 
\begin{eqnarray}
\Gamma _\sigma ^{\rm{eff}} (\omega ) &=& (1 + \alpha _V n_{d - \sigma } )^2 \nonumber \\
&\times& \pi \left[ {\left| {V_R } \right|^2 \rho _R^\sigma  (\omega ) + \left| {V_L } \right|^2 \rho _L^\sigma  (\omega )} \right],
\label{Gamaeff}
\end{eqnarray}   			
where $\rho _\alpha^\sigma  (\omega )$  is the density of states of $\alpha$ lead for spin $\sigma$. Within the wide band approximation the couplings become energy independent, $\Gamma_{\sigma }^{\rm{eff}}(\omega)=\Gamma_{\sigma }^{\rm{eff}}=\left(1+\alpha_V n_{d-\sigma}\right)^2 \Gamma$ .

The correlation parameter $B_\sigma$, appearing in Eq. \eqref{Sigmad}, can be written as \cite{Gorski-2016}:

\begin{eqnarray}
B_\sigma&=&- \sum\limits_{k \alpha} {2\Delta V_{k \alpha}}  \\
&\times&\int\limits_{ - \infty }^\infty  {\frac{1}{\pi }{\mathop{\rm Im}\nolimits} } \left( {\frac{{V_{k \alpha \sigma}^{\rm{eff}} }}{{\omega  - \varepsilon_{k \alpha} + \mu _\sigma}}\langle \langle d_\sigma;d_\sigma^{+}\rangle\rangle_\omega} \right)f(\omega )d\omega. \nonumber \label{BS}  
\end{eqnarray}
	  
The second order self-energy of the quantum dot, $\Sigma _{d\sigma }^{'}(\omega )$, is given by the expression \ref{sec.appendixB}: 
 \begin{eqnarray}
\Sigma _{d\sigma }^{'}(\omega ) = \frac{{\Sigma _{d\sigma }^{(2)} (\omega )}}{{1 + A_1 \Sigma _{d\sigma }^{(2)} (\omega )}},
\label{SigmaH}
\end{eqnarray} 	
where $\Sigma _{d\sigma }^{(2)} (\omega )$ is the second order self-energy
 \begin{eqnarray}
\Sigma _{d\sigma }^{(2)} (\omega ) = \frac{i}{{2\pi }}\int\limits_{ - \infty }^\infty  {dx} \frac{{\Sigma _{d\sigma }^{(2) > } (x) - \Sigma _{d\sigma }^{(2) < } (x)}}{{\omega  - x + i\eta }},
\label{Sigma2}
\end{eqnarray} 						
and
 \begin{eqnarray}
\Sigma _{d\sigma }^{(2) > {\rm  } < } (x) &=& U^2 { \frac{1}{(2\pi )^2 }} \\
&\times&\int\!\!\!\int {g_{d - \sigma }^{ < {\rm  } > } (y)g_{d - \sigma }^{ > {\rm  } < } (z)g_{d\sigma }^{ > {\rm  } < } (x - z + y)dydz}. \nonumber 
\label{Sigma2MW}
\end{eqnarray}

The functions $g_{d - \sigma }^{ < {\rm  } > }$ in Eq.~\eqref{Sigma2MW} are the lesser and the greater Green's functions, respectively. They can be calculated by the EOM method. To truncate the higher order GF we will use the Ng approximation \cite{Ng-1996}, which yields:
\begin{eqnarray}
g_{d\sigma }^ <  (x) =  - i2\pi f_{\rm{eff},\sigma } (x){\mathop{\rm Im}\nolimits} g_{d\sigma }^{\rm{HF}}(x),
\label{g0m}
\end{eqnarray}  
\begin{eqnarray}
g_{d\sigma }^ >  (x) = i2\pi (1 - f_{\rm{eff},\sigma } (x)){\mathop{\rm Im}\nolimits} g_{d\sigma }^{\rm{HF}} (x),
\label{gow}
\end{eqnarray}  
where the effective Fermi function is defined as:
 \begin{eqnarray}
f_{\rm{eff},\sigma } (x) = \frac{{\sum\limits_\alpha  {\Gamma _{\sigma \alpha }^{\rm{eff}} f(x - \mu _\alpha  )} }}{{\Gamma _\sigma ^{\rm{eff}} }}.
\label{fermi}
\end{eqnarray} 

The effective Green's function $g_{d\sigma }^{\rm{HF}} (x)$ is determined by the following one-body Hamiltonian
 \begin{eqnarray}
H_0  &=& \sum\limits_\sigma  {\varepsilon _{\rm{eff},\sigma} \hat n_{d\sigma} }  + \sum_{\substack{k\sigma  \\ \alpha  = L,R}} {(\varepsilon _{k\alpha }  - \mu _\alpha  )\hat n_{k\alpha \sigma } } \nonumber \\ 
	&+& \sum_{\substack{k\sigma  \\ \alpha  = L,R}} {\left( {V_{k\alpha\sigma}^{\rm{eff}} d_{\sigma}^{+} c_{k\alpha \sigma }  + h.c.} \right)} ,
\label{Hamiltonian0}
\end{eqnarray} 
where $\varepsilon _{\rm{eff},\sigma}$ is the effective quantum dot level selected in such a way that the  charge consistency between the one-body \eqref{Hamiltonian0} and the interacting problems \eqref{Hamiltonian} is achieved \cite{Potthoff-1997,Vecino-2003}. The another way of defining the effective quantum dot level is to assume that the Friedel sum rule should be fulfilled, what gives  $\varepsilon _{\rm{eff},\sigma}=\varepsilon _{d\sigma} +\Sigma_{d\sigma}(0)$ \cite{Kajueter-1996,Yamada-2011}, but this condition is strict for zero temperature only, and as was shown by Levy Yeyati and co-authors \cite{c57}, it is consistent with the constraint $\left\langle n_{0 d \sigma}\right\rangle=\left\langle n_{d \sigma}\right\rangle$. In our computations we will use this constraint.

\section{ The results}
In our numerical analysis we use the symmetrical coupling of the quantum dot with the left and right leads $\Gamma_R=\Gamma_L=\Gamma/2$. For the Coulomb interaction we use the value $U=5\Gamma$. The experimentally estimated values of $U$ are typically near $U/\Gamma=2$  \cite{Viel-2000}, 4.5 \cite{Kretinin-2011} up to  $U/\Gamma=7$ \cite{c51, Amasha-2005}. In addition, we assume that the average chemical potential of leads fulfills the relation $\mu_L + \mu_R=0$ and depends on the bias voltage $\mu_L - \mu_R=eV$. To analyze the influence of assisted coupling parameter $\alpha_V$ on the transport and magnetic properties of quantum dot we will use the negative values of this parameter. According to Eq.~\eqref{Gamaeff}, the negative value of $\alpha_V$  suppresses the effective coupling with leads. 

The assisted hopping term (correlated hopping) for the transition metal was estimated (first by Hubbard) to the value 0.5eV \cite{Hubbard-1963}. This estimation gives the value of assisted coupling parameter of the order of 0.3 to 1. For the superconductivity there was used a value of $\alpha_V$ of the order of 0.3 to 0.4 \cite{c27,c28}. To describe the transition between bond-order-wave and ferromagnetic ground states  Campbell et al. used $\alpha_V$ of the order of 0.1 to 0.2 \cite{Campbell-1990}. Vollhardt and co-authors \cite{c37,c38,c39} have shown, that the itinerant ferromagnetism can be obtained for $0<\alpha_V<1$. In our analysis we will use $\left|\alpha_V\right|\leq 0.5$. For such a value of assisted coupling parameter one can use the mean-field approximation.

\subsection{The case without magnetic field}
\begin{figure}[t]
\centering
\includegraphics[width=0.7\columnwidth]{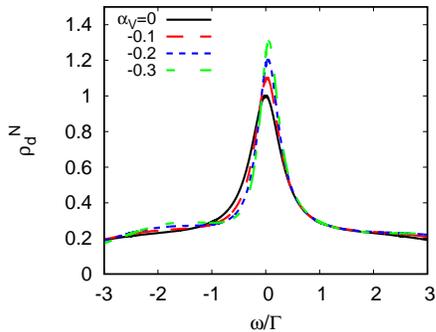} \caption{\label{fig:roalfa} 
The normalized spectral function of quantum dot in function of the assisted coupling parameter $\alpha_V$ for the quantum dot energy $\varepsilon_d=-U/2$ and $T=0$.}
\end{figure}

In Fig.~\ref{fig:roalfa} we present the spectral function of quantum dot as a function of the assisted coupling parameter $\alpha_V$ for the quantum dot energy  $\varepsilon_d=-U/2$. If we do not take into account the assisted coupling ($\alpha_V=0$), we obtain the symmetrical spectral function with Kondo peak localized near $\omega=0$. For the negative values of assisted coupling parameter $\alpha_V$ we observe the increase of spectral function for the Kondo peak. The maximum of spectral function shifts towards larger energies, $\omega>0$. Simultaneously, there is a decrease of the Kondo resonance width, because of the decrease of effective coupling with leads $\Gamma_\sigma^{\rm{eff}}$ (see Eq.~\eqref{Gamaeff}). 
 
\begin{figure}[tb]
\centering
\includegraphics[width=0.7\columnwidth]{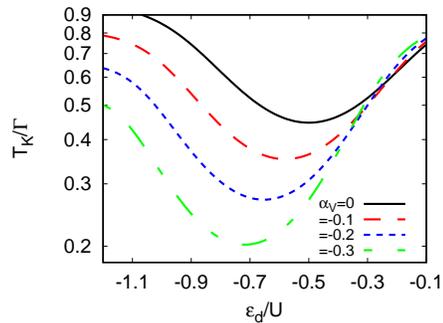} \caption{\label{fig:TK}  
The Kondo temperature (in logarithmic scale) as a function of the dot energy $\varepsilon_d$ for different values of the assisted coupling parameter $\alpha_V$.}
\end{figure}

Because the Kondo temperature $T_K$ depends on the Kondo resonance width \cite{Haldane-1978}, the decrease of effective coupling with leads $\Gamma_\sigma^{\rm{eff}}$, due to parameter $\alpha_V$, causes the decrease of $T_K$ value. 
In Fig.~\ref{fig:TK} we show the dependence of Kondo temperature $T_K$ (in logarithmic scale), estimated from the half-width at half maximum (HWHM) of the Kondo resonance at $T=0$, as a function of the quantum dot energy $\varepsilon_d$ for different values of assisted coupling parameter $\alpha_V$. 
For $\alpha_V=0$  the quantum dot energy dependence of the Kondo temperature is symmetrical with respect to $\varepsilon_d=-U/2$. Near $\varepsilon_d=-U/2$ the logarithmic behavior of the Kondo temperature $\rm{ln} \left(T_K\right)$ retains the parabolic character, but for $\varepsilon_d=-U$ and $\varepsilon_d=0$ the parabolic behavior is disturbed. The change of $\alpha_V$ value causes the decrease of Kondo temperature and additionally the minimum of the $T_K(\varepsilon_d)$  dependence shifts towards $\varepsilon_d<-U/2$. For $\alpha_V=-0.3$ the minimum Kondo temperature is shifted towards  $\varepsilon_d\approx-0.71U$. For this value of $\alpha_V$ parameter, the $T_K(\varepsilon_d)$ dependence is not symmetrical. 
Tooski et al. \cite{c45} shown that the Kondo temperature depends on the Kondo exchange coupling constant $J_K$. The assisted coupling parameter reduces the constant $J_K$ and shifts the minimum of $T_K(\varepsilon_d)$. The location of minimum of $T_K(\varepsilon_d)$ can be described by the relation $\varepsilon_d=-U/(1+\left|1+2\alpha_V\right|)$.

A similar asymmetry of the Kondo temperature was observed experimentally by Kretinin et al. \cite{Kretinin-2011} for the gate voltage dependence of Kondo temperature. As was shown in \cite{Kretinin-2011}, the dot energy depends linearly on the gate voltage. For the strong values of asymmetric hopping parameter we can see that the minimum of the Kondo temperature localizes near $(\varepsilon_d+U)/\Gamma=1$. A similar result was obtained experimentally for the single-molecule transistors based on transition-metal complexes \cite{c65}.
 
The assisted hopping effect also influences on the transport properties of the quantum dot. The electric current is calculated from the general formula for the current flowing through a region with interacting electrons \cite{Meir-1992,Jauho-1994}:  
\begin{eqnarray}
I &=& {\frac{2e} {\hbar} } {\sum\limits_\sigma {\int_{ - \infty }^\infty  {{\frac {\Gamma _{\sigma L}^{\rm{eff}} \Gamma _{\sigma R}^{\rm{eff}}} {\Gamma _\sigma ^{\rm{eff}} }}}}} \label{In}\\ 
& \times & \left[ {f(\omega  - \mu _L ) - f(\omega  - \mu _R )} \right]\rho _{d\sigma } (\omega )d\omega . \nonumber  
\end{eqnarray} 	 

\begin{figure}[t]
\centering
\includegraphics[width=0.7\columnwidth]{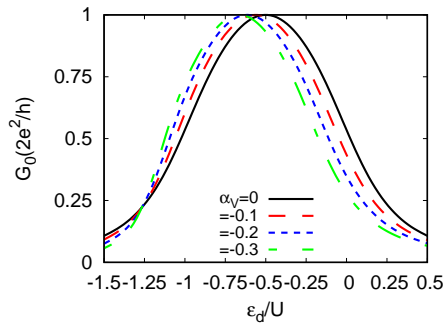} \caption{\label{fig:G0ed} Zero-bias differential conductance  $G_0=dI/dV(V=0)$ as a function of the dot energy $\varepsilon_d$ for different values of the assisted coupling parameter $\alpha_V$. }
\end{figure}

Equation \eqref{In} allows us to obtain the differential conductance $dI/dV$ by numerical differentiation of the total current. In Fig.~\ref{fig:G0ed} we show the differential conductance at zero bias voltage $G_0=dI/dV(V=0)$ as a function of the quantum dot energy $\varepsilon_d$ for different values of assisted coupling parameter $\alpha_V$ and for  $T=0$. For $\alpha_V=0$  the relation  $G_0(\varepsilon_d)$ is symmetrical around $\varepsilon_d=-U/2$, i.e. around the point for which the conductance approaches the unitary limit value $G_0=2e^2/h$ reported experimentally \cite{Kretinin-2011,Viel-2000}. The non-zero value of the assisted coupling parameter $\alpha_V$ causes that the zero-bias differential conductance still fulfills the unitary limit, but for the dot energy $\varepsilon_d<-U/2$. The assisted hopping also causes the symmetry breaking of $G_0(\varepsilon_d)$ characteristics, whereas if we present the dependence of zero-bias differential conductance as a function of the dot occupancy we obtain the symmetrical character of $G_0(n_{d\sigma})$. 

\begin{figure}[htb]
\centering
\includegraphics[width=0.7\columnwidth]{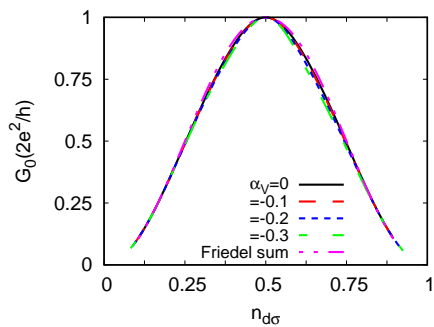} \caption{\label{fig:G0ns} The zero-bias differential conductance $G_0=dI/dV(V=0)$ as a function of the dot occupancy $n_{d\sigma}$ for different values of the assisted coupling parameter $\alpha_V$. For comparison, we show the theoretical dependence which is consistent with the Friedel sum rule. }
\end{figure}

In Fig.~\ref{fig:G0ns} we show the dependence  $G_0(n_{d\sigma})$ for different values of the assisted coupling parameter $\alpha_V$. For the comparison, we also  present the dependence of  $G_0(n_{d\sigma})=2e^2/h \sin^2(\pi n_{d\sigma})$ arising from the Friedel sum rule \cite{Kretinin-2011,Aligia-2006,Aligia-2004}. The obtained results show that for each value of the assisted coupling parameter we obtain the maximum of $G_0=2e^2/h$ for the half-filling dot occupancy ($n_{d\sigma}=0.5$). Additionally, our results agree very well with the theoretical dependence $G_0(n_{d\sigma})=2e^2/h \sin^2(\pi n_{d\sigma})$.

One of the fundamental issues related to the transport properties of quantum dots is the temperature dependence of zero-bias differential conductance $G(T)$. This dependence was examined experimentally \cite{Kretinin-2011,Viel-2000} and theoretically \cite{Ng-1988, Costi-2000, Costi-2001, Weymann-2013, Weymann-2011, Costi-2010}. In our analysis we will focus on the influence of the assisted coupling parameter $\alpha_V$ on the shape of $G(T)$ dependence. Meir, Hirose and Wingreen \cite{Meir-2002} proposed the occupancy dependent hybridization as a mechanism which explains the "0.7 anomaly" plateau of the zero-bias differential conductance $G(T)$ for the quantum point contacts.

\begin{figure}[t]
\centering
\includegraphics[width=1\columnwidth]{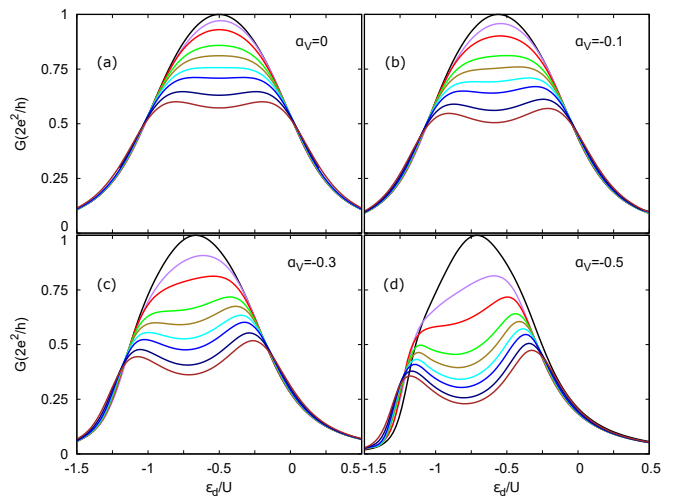} \caption{\label{fig:GedndkTalfa} 
The zero-bias differential conductance $G$ as a function of the dot energy $\varepsilon_d$  for different temperatures  $T/\Gamma=0$ (top), 0.03, 0.05, 0.08, 0.1, 0.125, 0.15, 0.2, 0.25 (bottom) and for $\alpha_V=0$ (a), $\alpha_V=-0.1$ (b), $\alpha_V=-0.3$ (c) and $\alpha_V=-0.5$ (d).
}
\end{figure}

\begin{figure}[th]
\centering
\includegraphics[width=0.6\columnwidth]{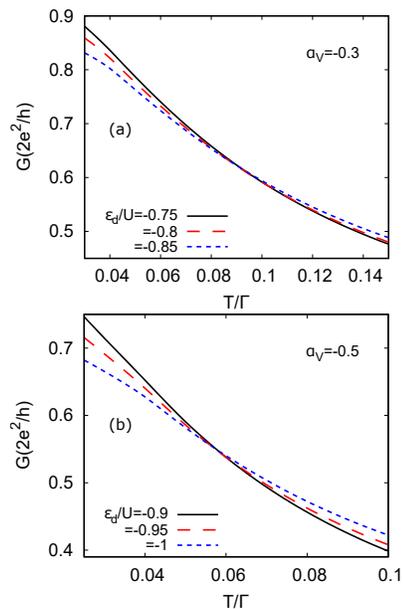} \caption{\label{fig:plateau} 
The zero-bias differential conductance $G$ as a function of the temperature  $T$ for different values of the dot energy $\varepsilon_d$ and for $\alpha_V=-0.3$ (a) and $\alpha_V=-0.5$ (b).
}
\end{figure}

Figure~\ref{fig:GedndkTalfa} presents the zero-bias differential conductance $G$ as a function of the dot energy $\varepsilon_d$ for different temperatures. When the temperature increases, the conductance becomes suppressed. For $\alpha_V=0$ (Fig.~\ref{fig:GedndkTalfa} (a)) the decrease of the zero-bias differential conductance has symmetrical character and is especially strong for $\varepsilon_d=-U/2$. For large values of temperature we observe the two Hubbard resonances separated by Coulomb blockade valley. 

For $\alpha_V\neq 0$ (Fig. \ref{fig:GedndkTalfa} (b)-(d)) we observe the asymmetrical dependence of $G(\varepsilon_d)$. 
For low values of $\alpha_V$ parameter the asymmetrical dependence of $G(\varepsilon_d)$ is weak. The increase of $\alpha_V$ parameter increases the asymmetry of conductance. At $\alpha_V=-0.3$ (Fig. \ref{fig:GedndkTalfa} (c)) the maximum of $G(T=0)$ is obtained for $\varepsilon_d=-0.66U$. For $\varepsilon_d<-0.66U$ the conductance drops faster with temperature than for $\varepsilon_d>-0.66U$. For high values of temperature the magnitude of the two Hubbard resonances is not identical. At $T\approx 0.085\Gamma$ there is a conductance plateau around $0.62(2e^2/h)$ (see Fig. \ref{fig:plateau} (a)). The increase of $\alpha_V$ parameter to the value of $-0.5$ (Fig. \ref{fig:GedndkTalfa} (d)) allows for obtaining a conductance plateau at $T\approx 0.06\Gamma$ (see Fig. \ref{fig:plateau} (b)). The location of this plateau depends on the $U/\Gamma$ factor. For low values of $U/\Gamma$ the plateau occurs near $0.5(2e^2/h)$, whereas for strong Coulomb correlation, the plateau is almost invisible. 
Lin et al.~\cite{c47}, which have used the numerical renormalization group method for $U/\Gamma\approx 10$, did not observe the conductance plateau even for strong assisted hopping interaction. If we assume that the plateau of conductance depends on the inter-site interactions (e.g. assisted hopping interaction), the too strong (dominant) Coulomb interaction, which destroys the Kondo effect with increasing temperature very fast, may prevent from occurring the plateau.

\begin{figure}[t]
\centering
\includegraphics[width=1\columnwidth]{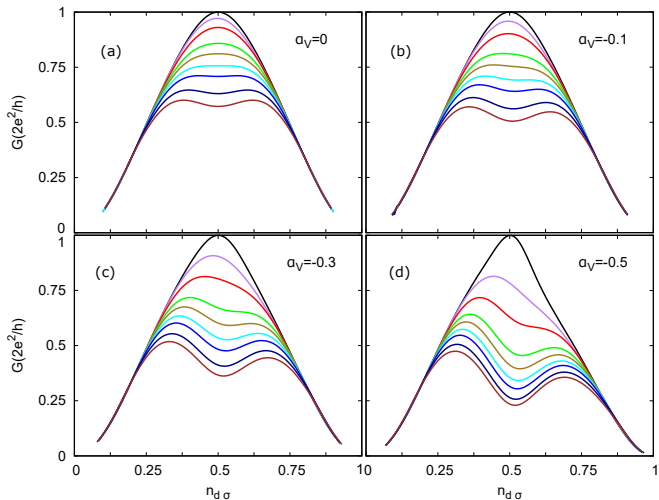} \caption{\label{fig:GndkTalfa} 
The zero-bias differential conductance $G$ as a function of the dot occupancy $n_{d\sigma}$ for different temperatures  $T/\Gamma=0$ (top), 0.03, 0.05, 0.08, 0.1, 0.125, 0.15, 0.2, 0.25 (bottom) and for $\alpha_V=0$ (a), $\alpha_V=-0.1$ (b), $\alpha_V=-0.3$ (c) and $\alpha_V=-0.5$ (d).
}
\end{figure}

Figure~\ref{fig:GndkTalfa} presents the zero-bias differential conductance $G$ as a function of the dot occupancy $n_{d\sigma}$ for different temperatures. At zero temperature, the conductance behavior is similar to the symmetrical dependence described by the $G_0(n_{d\sigma})=2e^2/h \sin^2(\pi n_{d\sigma})$ relation (see Fig.~\ref{fig:G0ns}). The increase of temperature for high values of $\alpha_V$ parameter causes the asymmetrical dependence of $G(n_d)$. 
The plateau of conductance is created for $n_{d\sigma}\approx0.5$, i.e. close to the symmetric point.

Our results can be compared with experimental data for InAs nanowire quantum dot obtained by Kretinin et al.~\cite{Kretinin-2011}. These authors analyzed the dependence of quantum dot conductance as a function of gate voltage, $V_g$, for different temperatures. In that work three regions of $G(V_g)$ were highlighted and labeled as III, IV and V Kondo valley. For the valley IV the coupling of quantum dot with leads is much weaker than for III and V valleys. The asymmetrical dependence of $G(\varepsilon_d)$, obtained by us for $\alpha_V=-0.3$ is similar to the $G(V_g)$ dependence for valley V. For valley III the $G(V_g)$ dependence asymmetry is much stronger. This behavior would require a higher values of assisted coupling parameter $\alpha_V$. For the IV Kondo valley the conductance plateau is not visible, while one can see the strong asymmetry of the Coulomb blockade peaks. This asymmetry is characteristic for the non-zero values of assisted coupling parameter $\alpha_V$ and for strong Coulomb interaction \cite{c47}. For the linear conductance of an AlGaAs/GaAs heterostructure \cite{Viel-2000} the asymmetry is visible near $V_g=-375mV$, i.e. in the region of stronger coupling with leads.

The plateau and asymmetry of conductance is also visible in experimental results for carbon nanotubes \cite{c63}. For large values of $\Gamma$ and $V_g=3.06 \div 3.1V$ the plateau of $G(V_g)$ appears at $T=200mK$, whereas for smaller values of $\Gamma$ $(V_g=-3.5 \div -3.46V$) the plateau of $G(V_g)$ is obtained at $T=75mK$. Additionally, the strong asymmetry for $G(V_g)$ dependence appears.

\begin{figure}[th]
\centering
\includegraphics[width=0.6\columnwidth]{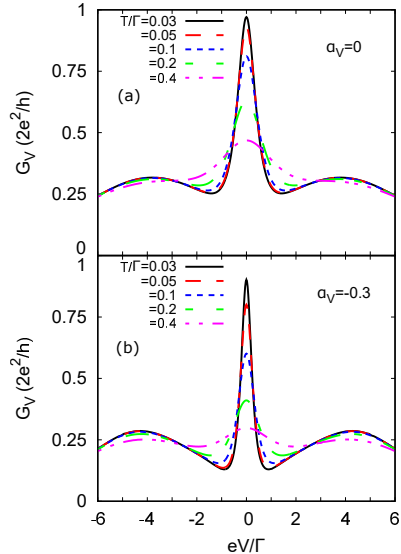} \caption{\label{fig:GV} 
Differential conductance $G_V=dI/dV$  as a function of the applied voltage between drain and source for different values of the temperature and assisted coupling parameter  $\alpha_V=0$ and $\varepsilon_d=-0.5U$ (a);  $\alpha_V=-0.3$ and $\varepsilon_d=-0.66U$ (b). 
}
\end{figure}

In Fig.~\ref{fig:GV} we show the differential conductance $G_V=dI/dV$ as a function of the applied voltage between drain and source, for different values of temperature. In the numerical computations we have used two sets of parameters: $\alpha_V=0$  and $\varepsilon_d=-0.5U$ in Fig.~\ref{fig:GV}(a);   $\alpha_V=-0.3$  and $\varepsilon_d=-0.66U$  in Fig.~\ref{fig:GV}(b), which correspond to the  maximum of linear conductance for the corresponding values of assisted coupling parameter. In both figures there are visible three maxima of differential conductance, zero-bias conductance anomaly and two Coulomb blockade peaks. With increasing temperature the magnitude of zero-bias conductance anomaly strongly decreases, while the magnitude of Coulomb blockade peaks changes slightly. In the case without correlated hybridization ($\alpha_V=0$), Fig.~\ref{fig:GV}(a), we obtain the larger width of zero-bias conductance anomaly than for $\alpha_V=-0.3$ (Fig.~\ref{fig:GV}(b)). The use of non-zero value of assisted coupling parameter does not cause the symmetry break of the differential conductance $G_V=dI/dV$. Analyzing the experimental results \cite{Kretinin-2011,Viel-2000}  one can also see the symmetry of differential conductance $G_V=dI/dV$.

\subsection{The effect of magnetic field}

Let us now consider the case when the external magnetic field, $\mathbf{B}$, is applied to the quantum dot. In the presence of $\mathbf{B}$ the spin-degeneracy is broken and we obtain two dot energies $\varepsilon_{d\uparrow}=\varepsilon_d-B$ and $\varepsilon_{d\downarrow}=\varepsilon_d+B$. 
	
\begin{figure}[tb]
\centering
\includegraphics[width=0.6\columnwidth]{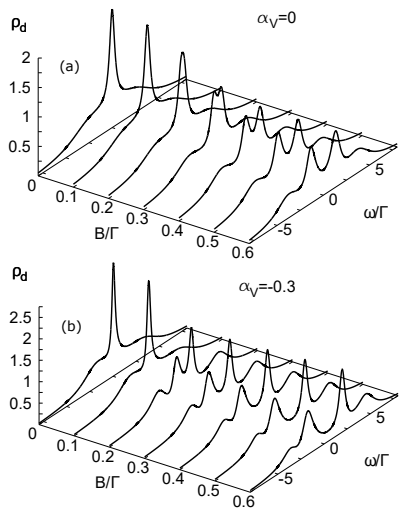} \caption{\label{fig:roB} 
The external magnetic field dependence of the normalized spectral function for $\varepsilon_d=-0.5U$, $T=0$ and two values of the assisted coupling parameter. The solid black line corresponds to the majority spin spectral function $\rho_d^\uparrow$ and the dashed red line corresponds to the minority spin spectral function $\rho_d^\downarrow$.
}
\end{figure}

\begin{figure}[tb]
\centering
\includegraphics[width=0.6\columnwidth]{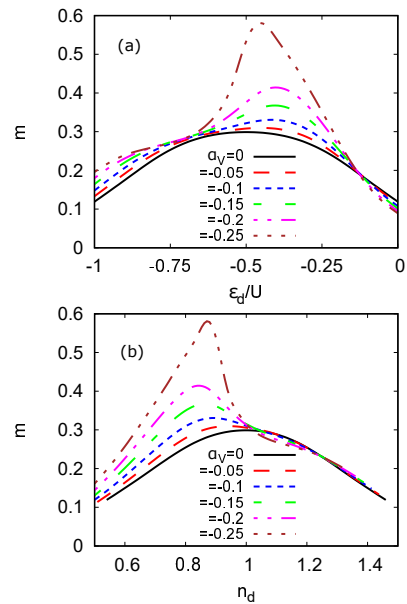} \caption{\label{fig:mednd} 
The magnetization as a function of dot energy $\varepsilon_d$ (a) and total occupancy $n_d$ (b) for different values of the assisted coupling parameter $\alpha_V$ and for $B=0.2\Gamma$.}
\end{figure}

In Fig.~\ref{fig:roB} we show the quantum dot normalized spectral function for different values of external magnetic field $\mathbf{B}$ and for $\alpha_V=0$ (Fig.~\ref{fig:roB}(a)) and $\alpha_V=-0.3$ (Fig.~\ref{fig:roB}(b)). The external magnetic field $\mathbf{B}$ causes the splitting of the Kondo peak. For low values of $\mathbf{B}$ the splitting is very weak and the total spectral function has one, central peak \cite{Kretinin-2011}. For high values of $\mathbf{B}$ we obtain the split of Kondo peak into two subresonances. The reduction of the total spectral function height occurs. At $\alpha_V=-0.3$ (Fig.~\ref{fig:roB}(b)) one can see that the splitting of Kondo peak is stronger than for $\alpha_V=0$. This leads to the stronger magnetization (see Fig.~\ref{fig:mednd}). Comparing Fig.~\ref{fig:roB}(a) and Fig.~\ref{fig:roB}(b) one can see that for $\varepsilon_d=-U/2$ and $\alpha_V=0$ the spectral functions are symmetrical, but for $\alpha_V=-0.3$ these functions are asymmetrical. 

The Kondo peak splitting under the influence of external magnetic field was reported in many previous works \cite{Meir-1993,Aligia-2006,Costi-2001,Moore-2000,Zitko-2009}. The external magnetic field causes also the splitting of the Kondo resonance in  $G_V=dI/dV$ as a function of the applied voltage between drain and source \cite{Meir-1993, Aligia-2006, c64}. In our analysis the influence of the assisted hopping effect on the magnetic properties of quantum dot will be more important. 

In Fig.~\ref{fig:mednd} we show the magnetization $m=\left\langle n_{d\uparrow}\right\rangle - \left\langle n_{d\downarrow}\right\rangle$  as a function of dot energy $\varepsilon_d$ (Fig.~\ref{fig:mednd}(a)) and total occupancy $n_d$ (Fig.~\ref{fig:mednd}(b)) for different values of the assisted coupling parameter $\alpha_V$ and for $B=0.2\Gamma$. For $\alpha_V=0$ (solid black line) we obtain the symmetrical dependence of $m(\varepsilon_d)$  and $m(n_d)$ with its maximum at $\varepsilon_d=-U/2$ (the particle-hole symmetry point $n_d=1$). In this case, when one moves away from the particle-hole symmetry point, one observes the decrease of magnetization. 

The more interesting characteristics of  $m(\varepsilon_d)$  and $m(n_d)$ is observed for the non-zero values of the assisted coupling parameter $\alpha_V$. The assisted hopping effect causes that the dependencies $m(\varepsilon_d)$ and $m(n_d)$ are asymmetric. For $\varepsilon_d>-U/2$ we observe the strong growth of magnetization with the change of $\alpha_V$. This behavior is due to fact, that the splitting of the dot level, generated by the external magnetic field, is additionally amplified by the correlation parameter $B_\sigma$ [given by Eq.~\eqref{BS}], which in the ferromagnetic case is spin dependent, $B_\uparrow \neq B_\downarrow$ \cite{Gorski-2016}. 

Additionally, for $\alpha_V<0$ the quantum dot density of states for Kondo peak increases (see Fig.~\ref{fig:roalfa})), causing the additional increase of magnetic moment \cite{Gorski-2016}. For $\varepsilon_d\approx0$ we observe the decrease of magnetization. For $\varepsilon_d<-U/2$ the magnetization enhancement caused by the assisted hopping effect is weak. In this region the quantum dot density of states for Kondo peak increases, but the correlation parameter $B_\sigma$ decreases the split of Kondo peak. The competition between these two effects causes that the magnetization depends mainly on the external magnetic field. In Fig.~\ref{fig:mednd}(b) we present the dependence of $m(n_d)$. For $n_d<1$ the magnetization is strongly enhanced by the assisted hopping effect, but for $n_d>1$ the magnetization is dependent on the external magnetic field $\mathbf{B}$ only. For small values of $\mathbf{B}$ the magnetization shows the Fermi-liquid behavior and grows linearly with field strength. For larger values of $\mathbf{B}$ we obtain the plateau of $m(B)$ corresponding to the saturated polarization. A similar dependence of magnetization on the external magnetic field was reported by Heyder and co-authors \cite{c62} for SIAM model and for Kondo quantum dot.

\begin{figure}[t]
\centering
\includegraphics[width=0.7\columnwidth]{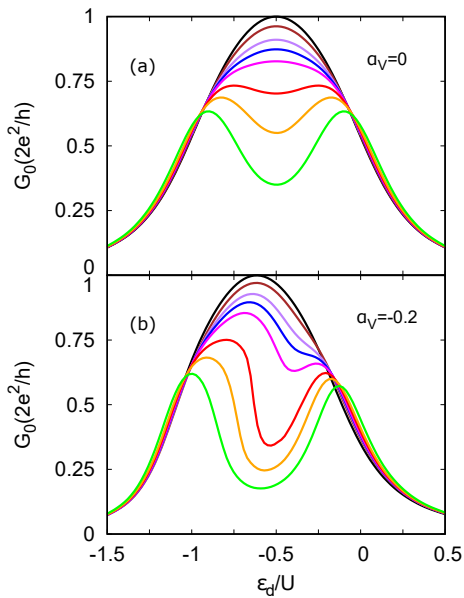} \caption{\label{fig:G0BB} 
The linear conductance $G_0$ as a function of the dot energy $\varepsilon_d$ at $\alpha=0$  (a) and $\alpha=-0.2$ (b), $T=0$and for different values of magnetic field strength $B/\Gamma=0$  (top), 0.1, 0.15, 0.175, 0.2, 0.25, 0.3, 0.4 (bottom).
}
\end{figure}

The dot energy ($\varepsilon_d$) dependence of the linear conductance for several values of magnetic field strength and for $T=0$ is presented in Fig.~\ref{fig:G0BB}. In Fig.~\ref{fig:G0BB}(a) we show the relation $G_0(\varepsilon_d)$ for $\alpha_V=0$. The increase of magnetic field causes strong decrease of the linear conductance in the Kondo regime. In the mixed valence regime this effect is much weaker, but in the the empty orbital regime we observe the weak increase of the linear conductance \cite{Costi-2001}. In this case this relation is symmetrical around $\varepsilon_d=-U/2$. The decrease of linear conductance with growing magnetic field is caused by the splitting of the Kondo peak \cite{Aligia-2006} and decrease of spectra density for the Fermi level (see Fig.~\ref{fig:roB}).

The non-zero value of assisted coupling parameter ($\alpha_V=-0.2$) causes that the decrease of linear conductance in the Kondo regime is much stronger (see Fig.~\ref{fig:G0BB}(b)) with respect to the results obtained for $\alpha_V=0$. For large values of magnetic field and for $\alpha_V=-0.2$ the minimum of linear conductance is achieved for the dot energy near $\varepsilon_d=-0.61U$. The relation $G_0(\varepsilon_d)$  has asymmetrical character. Stronger decrease of the linear conductance is observed for $\varepsilon_d>-0.61U$, i.e. in the region which corresponds to the strong magnetization (see Fig.~\ref{fig:mednd}(a)). As we have shown earlier, in this region the Kondo peak splitting effect assisted by the correlation parameter $B_\sigma$ is stronger. For $\varepsilon_d<-0.61U$  the reduction effect of the linear conductance with respect to the values obtained at $\alpha_V=0$ is much weaker. The asymmetry of $G_0(\varepsilon_d)$ for different values of external magnetic field was also obtained for Kondo quantum dot \cite{c62}. For appropriately selected value of magnetic field one can see the plateau for the linear conductance. This behavior is analogous to that observed as a function of the temperature.

\begin{figure}[tb]
\centering
\includegraphics[width=0.7\columnwidth]{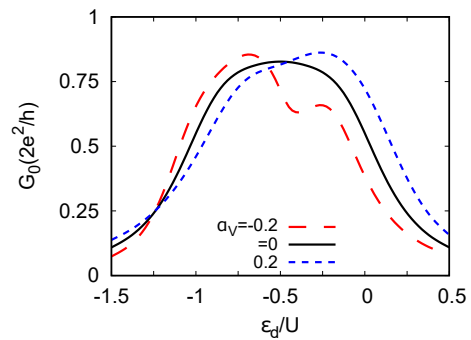} \caption{\label{fig:G0BBkalfa} 
The linear conductance $G_0$  as a function of the dot energy  $\varepsilon_d$ for $B/\Gamma=0.2$ and different values of assisted hopping parameter.
}
\end{figure}

For the system without external magnetic field the conductance $G_0(\varepsilon_d)$ for $\alpha_V<0$ can be mapped onto that  $\alpha_V>0$ by changing $\varepsilon_d$ to $-(\varepsilon_d+U)$ what can be expressed as $G_0(\alpha_V,\varepsilon_d)=G_0(-\alpha_V,-\varepsilon_d-U)$. The use of external magnetic field breaks this symmetry. In Fig.~\ref{fig:G0BBkalfa} we show the comparison of linear conductance $G_0$ as a function of the dot energy $\varepsilon_d$ for $\alpha_V=-0.2$  and $\alpha_V=0.2$. There is also presented the $G_0(\varepsilon_d)$ dependence without assisted hopping effect ($\alpha_V=0$). For negative value of the assisted hopping parameter we observe strong decrease of  the linear conductance $G_0$. This effect is highly visible in the Kondo regime, where for $\alpha_V<0$ the decrease of $G_0(\varepsilon_d)$ is connected to the strong increase of the magnetic moment, see Fig.~\ref{fig:mednd}(a). For $\varepsilon_d<-U$ the conductance is localized around the values corresponding to  $G_0(\varepsilon_d)$ for $\alpha_V=0$, whereas for $\varepsilon_d>0$ we obtain constant difference of $G_0(\varepsilon_d)$.

\section{Summary and conclusions}
In this paper we analyzed the influence of the assisted hopping process on the transport and magnetic properties of the quantum dot coupled with two metallic leads. Using the modified equation of motion approach and self-consistent perturbation theory we showed that the assisted hopping process can be responsible for the plateau in the linear conductance $G(\varepsilon_d)$ at different values of temperature. Similar plateau was observed for the quantum point contacts \cite{Meir-2002} and was reported experimentally for the nanowire quantum dot \cite{Kretinin-2011} and the carbon nanotube \cite{c63}. This process breaks the particle-hole symmetry of the zero-bias differential conductance, $G_0(\varepsilon_d)$. Additionally, the assisted hopping strongly affects the Kondo temperature and causes the Kondo peak asymmetry. 
 
In presence of the external field, the assisted hopping process enhances the spin polarization of electrons in quantum dot. This effect is related to the fact, that the splitting of the dot level, generated by the external magnetic field, is additionally amplified by the correlation parameter. The correlated hybridization also disturbs the symmetry of zero-bias conductance $G_0(\varepsilon_d)$  in the Kondo regime for small values of magnetic field. For appropriately selected values of the external field the appearance of of a plateau in the linear conductance $G_0(\varepsilon_d)$ is possible.

\section{Acknowledgments}
This work was done due to support from Faculty of Mathematics and Natural Sciences University of Rzesz\'{o}w within the project no.
WMP/GD-06/2017 and due to partially supported by the Center for Innovation and Transfer of Natural Sciences and Engineering Knowledge of Rzesz\'{o}w University.

\begin{widetext}
\appendix
\section{The modified equation of motion method}
\label{sec.appendixA}

The classical EOM method gives the wrong results in the particle-hole symmetric case because it causes the disappearance of the Kondo peak. 
In this method, one uses only one version of the equation of motion (with differentiation over the first time variable \eqref{EOM}).
As the result, we obtain an infinite hierarchy of equations of higher order Green's functions. Taking into account this fact, one should use the approximation that truncates this hierarchy and gives a closed set of equations which are self-consistently solvable. In this method, the differentiation over a second time variable \eqref{EOM2} is not used, but we have to pay attention that using only a second version of EOM \eqref{EOM2} gives identical results as in the case when we use the equation \eqref{EOM} only.
The modified equation of motion method, proposed by us, is based on the use of both EOM versions (\eqref{EOM} and \eqref{EOM2}). 
As the first step, we use the Eq. \eqref{EOM}. For the correlated system, we obtain the higher-order Green's functions, for which (in the second step) we use the second form of EOM \eqref{EOM2}.

Applying Eq. \eqref{EOM} to the function $\langle\langle d_\sigma;d_\sigma^\dag\rangle\rangle_\omega$ we obtain:
\begin{eqnarray}
\left( \omega-\varepsilon_d\right)\langle\langle d_\sigma;d_\sigma^\dag\rangle\rangle_\omega =
1+\sum_{k\alpha}V_{k\alpha}\langle\langle c_{k\alpha\sigma};d_\sigma^\dag\rangle\rangle_\omega
+ U\langle\langle\hat n_{d-\sigma}d_\sigma;d_\sigma^\dag  \rangle \rangle _\omega \nonumber \\
 +\sum_{k\alpha}\alpha_V V_{k\alpha}\langle\langle \hat n_{d-\sigma} c_{k\alpha\sigma};d_\sigma^\dag\rangle\rangle_\omega
+\sum_{k\alpha}\alpha_V V_{k\alpha}\langle\langle \left( d_{-\sigma}^\dag c_{k\alpha-\sigma} +c_{k\alpha-\sigma}^\dag d_{-\sigma} \right) d_\sigma;d_\sigma^\dag\rangle\rangle_\omega.  
\label{A1}
\end{eqnarray}

For the function $\langle\langle c_{k\alpha\sigma};d_\sigma^\dag\rangle\rangle_\omega$ we obtain the equation:
\begin{eqnarray}
(\omega-\varepsilon_{k\alpha\sigma}+\mu)\langle\langle c_{k\alpha\sigma};d_\sigma^\dag\rangle\rangle_\omega=V_{k\alpha}\langle\langle d_{\sigma};d_\sigma^\dag\rangle\rangle_\omega+
\alpha_V V_{k\alpha}\langle\langle\hat n_{d-\sigma}d_\sigma;d_\sigma^\dag\rangle\rangle_\omega
\label{A2}
\end{eqnarray}

Next, we will assume that the dominant energy scale is the Coulomb interaction. The assisted coupling between the $\alpha$ lead and the dot, $\alpha_V V_{k\alpha}$, will be much smaller than the Coulomb interaction, so for the Green's functions related to $\alpha_V V_{k\alpha}$ interaction we use the Hartree-Fock approximation, what causes that Eq. \eqref{A2} can be presented as:
\begin{eqnarray}
(\omega-\varepsilon_{k\sigma}+\mu)\langle\langle c_{k\sigma};d_\sigma^\dag\rangle\rangle_\omega=V_{k\alpha\sigma}^{\rm{eff}}\langle\langle d_{\sigma};d_\sigma^\dag\rangle\rangle_\omega
\label{A5}
\end{eqnarray}
where $V_{k\alpha \sigma }^{\rm{eff}}=V_{k\alpha } \left(1+\alpha_V n_{d-\sigma}\right)$ is the effective hybridization matrix element.

For the Green's functions related to $U$ we will use the irreducible Green's functions technique \cite{Ochoa-2014,Kuzemsky-2002}
\begin{eqnarray}
{}^{\rm{ir}}\langle\langle[A,H]_{-};B\rangle\rangle_\omega=\langle\langle[A,H]_{-}- z A;B\rangle\rangle_\omega
\label{A3}
\end{eqnarray}
where the $z$ constant represents the self-energy in the Hartree-Fock approximation. 

The characteristic property of the irreducible Green's functions ${}^{\rm{ir}}\langle\langle[A,H]_{-};B\rangle\rangle_\omega$ is that they cannot be
reduced to the lower-order Green's functions by any kind of decoupling.

Using the irreducible Green's functions method we obtain the equations:
\begin{eqnarray}
\left( \omega-\varepsilon_d-U n_{d-\sigma}- B_\sigma \right)\langle\langle d_\sigma;d_\sigma^\dag\rangle\rangle_\omega\;= 
1+\sum_{k}V_{k\alpha\sigma}^{\rm{eff}}\langle\langle c_{k\sigma};d_\sigma^\dag\rangle\rangle_\omega+ U {}^{\rm{ir}}\langle\langle\hat n_{d-\sigma}d_\sigma;d_\sigma^\dag  \rangle \rangle _\omega  
\label{A4}
\end{eqnarray}
where $B_\sigma=\sum_{k}\alpha_V V_{k\alpha}(\langle d_{-\sigma}^\dag c_{k-\sigma} \rangle+\langle c_{k-\sigma}^\dag d_{-\sigma} \rangle)$ is the correlation parameter.

To obtain the ${}^{ir}\langle\langle\hat n_{d-\sigma}d_\sigma;d_\sigma^\dag\rangle\rangle_\omega$ function, we will use the differentiation with respect to the
second time variable (see Eq. \eqref{EOM2}). The method of differentiation with respect to the first and second time variable was widely used by Tserkovnikov \cite{Tserkovnikov-1981,Tserkovnikov-1999} and Kuzemsky \cite{Kuzemsky-2002}. Using Eq. \eqref{EOM2} we obtain the equation:
\begin{eqnarray}
\left( \omega-\varepsilon_d-U n_{d-\sigma}- B_\sigma \right) {}^{\rm{ir}}\langle\langle\hat n_{d-\sigma}d_\sigma;d_\sigma^\dag\rangle\rangle_\omega \;= 
\sum_{k}V_{k\alpha}^{\rm{eff}}  {}^{\rm{ir}}\langle\langle \hat n_{d-\sigma}d_\sigma;c_{k\sigma}^\dag\rangle\rangle_\omega
+ U {}^{\rm{ir}}\langle\langle\hat n_{d-\sigma}d_\sigma;\hat n_{d-\sigma} d_\sigma^\dag\rangle \rangle _\omega^{\rm{ir}} 
\label{A6}
\end{eqnarray}

For the function ${}^{\rm{ir}}\langle\langle \hat n_{d-\sigma}d_\sigma;c_{k\alpha\sigma}^\dag\rangle\rangle_\omega$, after using Eq. \eqref{EOM2} and the Hartree-Fock approximation, we obtain:
\begin{eqnarray}
(\omega-\varepsilon_{k\alpha\sigma}+\mu) {}^{\rm{ir}}\langle\langle\hat n_{d-\sigma}d_\sigma;c_{k\alpha\sigma}^\dag\rangle\rangle_\omega=V_{k\alpha}^{\rm{eff}} {}^{\rm{ir}}\langle\langle\hat n_{d-\sigma}d_\sigma;d_\sigma^\dag\rangle\rangle_\omega
\label{A7}
\end{eqnarray}

Using Eq. \eqref{A7} in \eqref{A6} we obtain: 
\begin{eqnarray}
\left( \omega-\varepsilon_d-U n_{d-\sigma}- B_\sigma -i\Gamma _\sigma ^{\rm{eff}} (\omega ) \right) {}^{\rm{ir}}\langle\langle\hat n_{d-\sigma}d_\sigma;d_\sigma^\dag\rangle\rangle_\omega = 
=U {}^{\rm{ir}}\langle\langle\hat n_{d-\sigma}d_\sigma;\hat n_{d-\sigma} d_\sigma^\dag\rangle \rangle _\omega^{\rm{ir}} 
\label{A8}
\end{eqnarray}
where $i\Gamma _\sigma ^{\rm{eff}} (\omega )=\sum_{k\alpha} \frac{\left|V_{k\alpha\sigma}^{\rm{eff}}\right|^2}{\omega-\varepsilon_{k\alpha\sigma}+\mu}$.

Using the symbols: 
\begin{eqnarray}
G_{d \sigma \rm{HF}}^{-1}(\omega)= \omega-\varepsilon_d-U n_{d-\sigma}- B_\sigma -i\Gamma _\sigma ^{\rm{eff}} (\omega )
\label{A9}
\end{eqnarray}
and
\begin{eqnarray}
F_{d\sigma}(\omega)={}^{\rm{ir}}\langle\langle\hat n_{d-\sigma}d_\sigma;\hat n_{d-\sigma} d_\sigma^\dag\rangle \rangle _\omega^{\rm{ir}} 
=\langle\langle(\hat n_{d-\sigma}- n_{d-\sigma})d_\sigma;(\hat n_{d-\sigma}- n_{d-\sigma}) d_\sigma^\dag\rangle \rangle _\omega
\label{A10}
\end{eqnarray}
with the Eqs. \eqref{A4} and \eqref{A8} we obtain the relations for the one-particle Green function of the quantum dot:
\begin{eqnarray}
\langle \langle d_\sigma  ;d_\sigma ^\dag  \rangle \rangle _\omega   =\frac {1} {\omega  - \varepsilon _d - B_\sigma - i\Gamma_\sigma^{\rm{eff}}(\omega )   - \Sigma _{d\sigma } (\omega )}
\label{A11}
\end{eqnarray}
where the self-energy for Coulomb interaction is equal to
\begin{eqnarray}
\Sigma _{d\sigma } (\omega ) = U\langle {n_{d-\sigma } }\rangle  + \frac{U^2 F_{d\sigma } (\omega )} {1 + U^2 F_{d\sigma } (\omega )G_{d\sigma \rm{HF} } (\omega )}.
\label{A12}
\end{eqnarray}
The obtained relation for the one-particle Green function of a quantum dot is modified by the assisted coupling of the two correlated effects: the spin dependent effective hybridization function and the spin dependent shift of the quantum dot energy. The self-energy for the Coulomb interaction is a sum of the Hartree-Fock part and the higher order part depending on the irreducible function $F_{d\sigma}(\omega)$. One should be mentioned that for $F_{d\sigma}(\omega)$ we use the functional form of the higher-order irreducible Green functions (see Appendix \ref{sec.appendixB}).

\section{The self-energy of correlated quantum dot}
\label{sec.appendixB}

The higher-order part of the self-energy for Coulomb interaction, given by Eq. \eqref{A12}, contains the complex prefactor $\left[ {1 + U^2 F_{d\sigma } (\omega )G_{d\sigma \rm{HF}} (\omega )} \right]^{ - 1}$, which for high values of $U$ does not allow for self-consistent solution at some energies $\omega$. In order to avoid this error we replaced the Green function $G_{d\sigma \rm{HF}} (\omega )$ by the parameter $A_1$, which is chosen in such a way to reproduce the exact result in the atomic limit \cite{Potthoff-1997,Kajueter-1996} 
\begin{eqnarray}
\Sigma _{d\sigma } (\omega ) = U\langle {n_{d-\sigma } }\rangle  + \frac{U^2 F_{d\sigma } (\omega )} {1 + A_1 U^2 F_{d\sigma } (\omega )}.
\label{B0}
\end{eqnarray}

To obtain the irreducible function $F_{d\sigma}(\omega)$ we have using the spectral theorem:
\begin{eqnarray}
{}^{\rm{ir}}\langle\langle\hat n_{d-\sigma}d_\sigma;\hat n_{d-\sigma} d_\sigma^\dag\rangle \rangle _\omega^{\rm{ir}}&=& 
\frac{1}{2\pi} \int\limits_{ - \infty }^\infty{\frac{\left[\exp{(\beta\omega')}+1\right]d\omega'}{\omega-\omega'}}\nonumber \\
&\times&\int\limits_{ - \infty }^\infty{\exp{(i\omega' t)} {}^{\rm{ir}}\langle \left[d_{-\sigma}^\dag(t)d_{-\sigma}(t)d_{\sigma}^\dag(t),d_{-\sigma}^\dag(0)d_{-\sigma}(0)d_{\sigma}(0)\right]_+\rangle^{\rm{ir}}dt}
\label{B1}
\end{eqnarray}

For the irreducible correlate function ${}^{\rm{ir}}\langle \left[d_{-\sigma}^\dag(t)d_{-\sigma}(t)d_{\sigma}^\dag(t),d_{-\sigma}^\dag(0)d_{-\sigma}(0)d_{\sigma}(0)\right]_+\rangle^{\rm{ir}}$ we will use the decouple scheme:
\begin{eqnarray}
{}^{\rm{ir}}\langle \left[d_{-\sigma}^\dag(t)d_{-\sigma}(t)d_{\sigma}^\dag(t),d_{-\sigma}^\dag(0)d_{-\sigma}(0)d_{\sigma}(0)\right]_+\rangle^{\rm{ir}}= 
\langle d_{-\sigma}^\dag(t)d_{-\sigma}(0)\rangle\langle d_{-\sigma}(t)d_{-\sigma}^\dag(0)\rangle\langle d_{\sigma}^\dag(t)d_{\sigma}(0)\rangle\nonumber \\
-\langle d_{-\sigma}(0) d_{-\sigma}^\dag(t)\rangle\langle d_{-\sigma}^\dag(0)d_{-\sigma}(t)\rangle\langle d_{\sigma}(0)d_{\sigma}^\dag(t)\rangle
\label{B2}
\end{eqnarray}
where the two-operator correlation functions with the same time variable are omitted (see \eqref{A3}).

For the two-operator time correlation functions we use 
\begin{eqnarray}
\langle A^\dag(0) B(t)\rangle=\frac{1}{\hbar}\int\limits_{ - \infty }^\infty{\frac{S_{AB}(\omega')d\omega'}{\exp{(\beta\omega')}+1}\exp\left(-\frac{i}{\hbar}\omega't\right)}
\label{B3}
\end{eqnarray}
and
\begin{eqnarray}
\langle B^\dag(t) A(0)\rangle=\frac{1}{\hbar}\int\limits_{ - \infty }^\infty{\frac{S_{AB}(\omega')\exp{(\beta\omega')}d\omega'}{\exp{(\beta\omega')}+1}\exp{\left(-\frac{i}{\hbar}\omega't\right)}}
\label{B4}
\end{eqnarray}
where $S_{AB}(\omega')=-\frac{1}{\pi}{\rm Im}\langle\langle B,A \rangle\rangle_{\omega'}$.

\begin{eqnarray}
F_{d\sigma } (\omega ) = \frac{i}{2\pi}\int\limits_{ - \infty }^\infty  {dx} \frac{F_{d\sigma }^ >  (x) - F_{d\sigma }^ <  (x)} {\omega  - x + i0^ +  },
\label{B5}
\end{eqnarray}
where 
\begin{eqnarray}
F_{d\sigma }^{ > {\rm  } < } (x) \approx \frac{1} {(2\pi )^2 }\int\!\!\!\int {g_{d - \sigma }^{ < {\rm  } > } (y)g_{d - \sigma }^{ > {\rm  } < } (z)g_{d\sigma }^{ > {\rm  } < } (x - z + y)dydz} 
\label{B6}
\end{eqnarray}
The $g_{d\sigma }^{<{\rm}>}(x)$ functions in Eq.~\eqref{B6} are the effective lesser $g_{d\sigma }^<(x)=- i2\pi f_{\rm{eff}} (x) {\rm Im} g_{d\sigma }^{\rm{eff}} (x)$ and greater $g_{d\sigma }^ >  (x) = i2\pi \left[ {1 - f_{\rm{eff}} (x)} \right] {\rm Im} g_{d\sigma }^{\rm{eff}} (x)$ Green's functions.
\end{widetext}

\end{document}